\documentclass[a4paper,11pt,leqno]{amsart}

\usepackage[utf8]{inputenc}
\usepackage{amsmath}
\usepackage{amsfonts}
\usepackage{amssymb}
\usepackage[left=2cm,top=3cm,right=2.5cm,bottom=2.5cm]{geometry}

\newcommand\tint{{\textstyle \int}}

\begin{document}

\title[Two-component integrable generalizations of  Burgers ...]
{Two-component integrable generalizations of Burgers equations with 
nondiagonal linearity}

\author[D. Talati]{Daryoush Talati}
\address{Department of Engineering Physics, Ankara University 
06100 Tando\u{g}an, Ankara}
\email{talati@eng.ankara.edu.tr,Daryoush.talati@gmail.com}
\author[R. Turhan]{Ref\.{i}k Turhan}
\address{Department of Engineering Physics, Ankara University 
06100 Tando\u{g}an, Ankara}
\email{turhan@eng.ankara.edu.tr}

\newtheorem{theorem}{Theorem}[section]
\newtheorem{proposition}{Proposition}
\newtheorem{remark}{Remark}
\begin{abstract}
Two-component second and third-order Burgers type systems with 
nondiagonal constant matrix of leading order  terms are 
classified for higher symmetries. New symmetry integrable systems with
their master symmetries are obtained. Some third order systems are observed 
to possess conservation laws. Bi-Poisson structures of systems possessing 
conservation laws are given. 
\end{abstract}

\maketitle

\section{Introduction}

Systematic classification of multi-component integrable equations is initiated 
by Mikhailov, Shabat and Yamilov in \cite{MSY87}. They completely classified 
second order, two-component systems of form 
\begin{equation*}
\frac{d{\bf u}}{dt}={\bf A}({\bf u}){\bf u}_{xx}+{\bf F}({\bf u},{\bf u}_x),\;\;
\det({\bf A}({\bf u}))\neq 0,\;\;
{\bf u}=\left(\!\!\begin{array}{c} u \\ v \end{array}\!\!\right),
\end{equation*}
possessing infinitely many conservation laws. 
One of their extensive results was the fact that all the integrable cases 
of the class considered were those systems that can be written by generalized 
contact transformations in the form in which the coefficient matrix of 
leading order terms ${\bf A}={\bf J}_2^{(-1)}$, where ${\bf J}_2^{(\lambda)}$ 
is the second of the three constant matrices
\begin{equation*}
{\bf J}_1=\left(
\begin{array}{cc}
1 & 0 \\
0 & 1
\end{array}
\right);\,
{\bf J}_2^{(\lambda)}=
\left(
\begin{array}{cc}
1 & 0 \\
0 & \lambda
\end{array}
\right),\lambda\neq1;\,
{\bf J}_3^{(a,\epsilon)}=
\left(
\begin{array}{cc}
a & \epsilon \\
0 & a
\end{array}
\right),\epsilon \ne 0.
\end{equation*}
Up to an overall factor which can be absorbed by the evolution parameter $t$, 
and permutation of rows which corresponds to rewriting the system with 
$u$ and $v$ interchanged, the above matrices, which are deduced from the 
canonical $2 \times 2$ Jordan matrices, exhaust all possibilities for the 
constant coefficient matrices of linear (in leading $x$-derivative order $M$) 
terms of any quasilinear two-component system
\begin{equation*}
\frac{d{\bf u}}{dt}={\bf A}{\bf u}_{M}+lower\;\;order\;\; terms,
\end{equation*} 
under the linear change of dependent variables 
(similarity transformations of constant matrix ${\bf A}$). 
Also, up to a rescaling of $t$,  $a=0$ and $a=1$ suffice for systems with 
${\bf A}={\bf J}_3^{(a,\epsilon)}$, 
and nonzero $\epsilon$ can be adjusted to any desired  nonzero value by a 
rescaling of $v$. So for quasilinear systems with nondiagonal linearity, 
the constant matrix ${\bf A}$ can be chosen, without loss of generality, 
to be ${\bf J}_3^{(1,1)}$ and  ${\bf J}_3^{(0,1)}$ 
to represent  the nodegenerate and the degenerate cases respectively.

Multi-component generalizations of second order 
\[\frac{du}{dt}=B_2[u]=u_{xx}+2uu_{x}\,,\] and third order 
\[\frac{du}{dt}=B_3[u]=u_{xxx} + 3uu_{xx}  + 3u^{2}u_{x} + 3u_{x}^{2}\,,\]
scalar Burgers equations have been the subject of various systematic 
integrability classifications.
Among these, those rewieved in \cite{MSS91} and 
the two-component cases of Svinolupov \cite{Svi89}  
were classifications with  assumption of matrix ${\bf A}={\bf J}_1$. 
In the classifications by Sanders and Wang \cite{SanWa}, 
and two-component cases of Tsuchida and Wolf \cite{TW} 
the case ${\bf A}={\bf J}_2^{(\lambda)}$ with arbitrary nonzero $\lambda$ 
are classified. 
Burgers type systems classified in \cite{Four00} were systems with matrix 
${\bf A}$ similar to ${\bf J}_1$ and ${\bf J}_2^{(0)}$.
Here we mention papers pertaining to multi-component generalizations 
of Burgers equation only. For the other integrable systems and their properties, 
we refer to review papers \cite{MSS91,ASY00,MNW09} and the references therein.

So far the only known integrable system with nondiagonal ${\bf A}$  
is the Karasu(Kalkanl{\i}) system
\begin{equation}\label{aysys}
\left\{
\begin{array}{l} 
\frac{dw}{dt}=w_{xxx} +z_{xxx} + 2w_{x}z+2wz_{x}\,, \\
\frac{dz}{dt}=z_{xxx} -9ww_{x} + 6zw_{x} + 3wz_{x} + 2zz_{x} \,,
\end{array}\right . 
\end{equation}
obtained in Painlav\'{e} classification \cite{AyK}. A recursion 
operator is given for the system in \cite{AAS} and its bi-Poisson formulation 
is given by Sergyeyev in \cite{ArS}. System~\eqref{aysys} is related to 
the system 
\begin{equation}\label{taysys}
\left\{
\begin{array}{l}
\frac{du}{dt}=u_{xxx} + 3uu_{xx}  + 3u^{2}u_{x} + 3u_{x}^{2} 
+ v_{xx} + 2vu_{x} \,,\\
\frac{dv}{dt}=v_{xxx} -3uv_{xx} + 6vu_{xx} + 3u_{x}v_{x} 
+ 3u^{2}v_{x} + 2vv_{x}\,,
\end{array}\right . 
\end{equation}
by the noninvertible Miura transformation $w=u_{x}$, $z=v+\frac{3}{2}u^{2}$. 
The reduction $v=0$ of the transformed system~\eqref{taysys} is nothing but 
the scalar third-order Burgers equation $\frac{du}{dt}=B_3[u]$. 

As a general rule, as the number of components and/or the order of a system 
increase, the computations needed for a higher symmetry rapidly goes beyond the 
available computing power. In such a situation it is not possible to consider 
a class of systems in its full generality. 
One of fruitful approaches to specify a rather restricted class of systems 
is to consider systems made up of only polynomial terms selected according 
to a certain weighting scheme. This approach is based on the observation 
that the vast majority of known integrable equations can be written in 
polynomial form in which they are homogeneous in a certain weighting scheme 
defined by invariance under scaling 
\begin{equation*}
(x,t,u,v)\rightarrow (a^{-1}x,a^{-\mu}t,a^{\lambda_1}u,a^{\lambda_2}v),
\quad a>0\,(\in \mathbb{R}).
\end{equation*}
Such a system is said to be $(\lambda_1,\lambda_2)$-homogeneous of weight 
$\mu$. In this terminology, system~\eqref{aysys} is $(2,2)$-homogeneous of 
weight 3 and its transformed form \eqref{taysys} is $(1,2)$-homogeneous of 
weight 3.

Here we classify $(1,1)$-homogeneous systems which are left unclassified in 
terms of the matrix of leading order terms in \cite{SanWa,TW,Four00}. 
I.e. we classify $(1,1)$-homogeneous of weight 2 and 3 class of (Bugers type) 
systems having nondiagonal constant coefficient matrix of leading order terms 
${\bf A}={\bf J}_3^{(a,\epsilon)}$, to admit a symmetry from the class of 
$(1,1)$-homogeneous of weight 1 and/or 2 higher than that of the class itself. 
We give complete list of systems having nondegenerate 
${\bf J}_3^{(a,\epsilon)}$, i.e. the case with $a\neq0$ and discuss some 
degenerate $(a=0)$ cases that arise as certain limits of nondegenerate cases 
obtained.

Two  systems of $N$-component evolutionary partial differential equations, 
or shortly systems, with characteristis ${\bf P}$ and ${\bf Q}$
\begin{equation}\label{evsys}
\begin{array}{l}
\frac{d{\bf u}}{dt}={\bf P}[{\bf u}]\,, \;\; 
\frac{d{\bf u}}{d\tau}={\bf Q}[{\bf u}]
\end{array}
\end{equation}
are said to be \emph{compatible} if their flows commute 
\begin{equation}\label{symcon}
\begin{array}{l}
\frac{d}{dt}\frac{d{\bf u}}{d\tau}=\frac{d}{d\tau}\frac{d{\bf u}}{dt}.
\end{array}
\end{equation}
Here, ${\bf u}=(u^1,u^2,\cdots,u^N)^{T}$, and the dependent variables $u^i$, 
$i=1,2,3,\cdots ,N$ are assumed to be dependent on independent variable $x$ 
and evolution parameter $t$.  ${\bf P}[{\bf u}]$ means that 
the components $P^i$ of ${\bf P}$ are differential functions depending smoothly 
on the independent ``spatial" variable $x$, dependent variables $u^i$ and 
finitely many $x$-derivatives of $u^i$ which will be denoted as 
$u^i_k=\frac{\partial^k u^i}{\partial x^k}$. 
For small values of $k\leq3$, we often use $u^i_x,u^i_{xx},u^i_{xxx}$ instead 
of $u^i_1,u^i_{2},u^i_{3}$. The highest order of $x$-derivative of $u^i$
in a differential function is referred as the order of the differential 
function.

Using chain rule in the left hand side of compatibility 
condition~\eqref{symcon} one gets
\begin{equation*}
\frac{d}{dt}{\bf Q}=\sum_
{\begin{subarray}{c} 1\leq i\leq N,\\k\in\mathbb{Z}_{\geq 0}\end{subarray}} 
D^k(P^i)\frac{\partial {\bf Q}}{\partial u^i_k}=X_{\bf P}{\bf Q} 
\end{equation*} 
where
\begin{equation*}
D=\frac{\partial}{\partial x}+\sum_
{\begin{subarray}{c} 1\leq i\leq N,\\k\in\mathbb{Z}_{\geq 0} \end{subarray}} 
u^i_{k+1}\frac{\partial}{\partial u^i_{k}}, 
\end{equation*}
is the total $x$-derivative and
\begin{equation*}
X_{\bf P}=\sum_
{\begin{subarray}{c} 1\leq i\leq N,\\k\in\mathbb{Z}_{\geq 0} \end{subarray}} 
D^k(P^i)\frac{\partial }{\partial u^i_k},
\end{equation*}
is the \emph{evolutionary vector field} with characteristic ${\bf P}$. 
Evolutionary vector fields with at most first order characteristics generate 
point transformations. Those with higher order characteristics are referred 
as generalized evolutionary vector fields.

By a similar identification in the right hand side,  
compatibility condition \eqref{symcon} reads
\begin{equation*}
\frac{d}{dt}\frac{d}{d\tau}{\bf u}-\frac{d}{d\tau}\frac{d}{dt}{\bf u}=
[\frac{d}{dt},\frac{d}{d\tau}]{\bf u}=
X_{\bf P}{\bf Q}-X_{\bf Q}{\bf P}=[X_{\bf P},X_{\bf Q}]=
X_{[X_{\bf P},X_{\bf Q}]}=0.
\end{equation*}
Therefore evolutionary systems in \eqref{evsys} are compatible if 
(and only if \cite{Olv93}) the Lie bracket $X_{[X_{\bf P},X_{\bf Q}]}$ 
of their corresponding  evolutionary vector fields vanishes.
An evolutionary vector field commuting with 
the characteristic of a system is defined as a \emph{symmetry} of the system.
If a (dispersive) system possess infinitely many symmetries 
(of arbitrarily higher orders) then the system is defined to be 
\emph{symmetry integrable}. A collection of mutually commuting evolutionary 
vector fields constitute an abelian Lie algebra which is also referred  
as symmetry algebra.

A particular system which is determined to possess a higher order symmetry 
is a natural subject of further investigations for structures like recursion 
operator \cite{Olv93} or master symmetry \cite{FoFu81,FiFo02} existence 
of which indicate existence of infinitely many higher symmetries. Because by 
definition a recursion operator maps a symmetry to another symmetry of a system 
and master symmetries do the same by their adjoint action on a given symmetry.

Some symmetry integrable systems can be included in an infinite hierarchy of  
compatible systems each of which is a conservation law
\begin{equation*}
\frac{d{\bf u}}{dt_n}={\bf P}_n[{\bf u}]=
{\mathcal H}{\bf \delta}\tint h_{n+1}[{\bf u}]dx
={\mathcal K}\delta\tint h_{n}[{\bf u}]dx,\; n=-1,0,1,2,\cdots,
\end{equation*}
written by two compatible Poisson structues ${\mathcal H}$ and ${\mathcal K}$,
 each, and sum of which are skew-adjoint operators satisfying Jacobi identity. 
Here, $\delta$ denotes the variational derivative with components 
$(\delta)_i=\frac{\delta}{\delta u^i}=\sum_{k\in \mathbb{Z}_\geq0} 
(-D)^k\frac{\partial}{\partial u^i_k}$, and $h_n$ 
are the Hamiltonian functionals in involution 
$\{\tint h_n,\tint h_m\}_{{\mathcal H}}=
\tint\delta h_n\cdot{\mathcal H}\delta h_mdx=0
=\tint \delta h_n\cdot{\mathcal K}\delta h_mdx=
\{\tint h_n,\tint h_m\}_{{\mathcal K}}$ 
with respect to both Poisson brackets corresponding to each Poisson structure.
Construction of such infinite hierarchy of compatible systems and the 
corresponding Hamiltonian functionals 
in involution, i.e. the Lenard-Magri scheme, in the case of local Poisson 
structures is rigorously addressed in \cite{BDSK09} for single-sided and 
in \cite{DSKT14,DSKT15} for multi-sided Lenard-Magri schemes.  
The case of nonlocal Poisson structures 
is treated in \cite{DSK13}, where a completely algorithmic criteria 
(i.e. without needing any user intervention) is given for checking whether 
a nonlocal skew-adjoint operator ${\mathcal K}$ which is a ratio of 
local operators ${\mathcal A}$ and ${\mathcal B}$ as
${\mathcal K}={\mathcal A}({\mathcal B})^{-1}$ satisfies the 
Jacobi identity or not.  
We do not go any further details of this broad subject here but refer 
to the mentioned references and to the book \cite{Dorf93} by Dorfman 
for a survey of early results. 

We  present results of our computer classification of $(1,1)$-homogeneous 
of weight 2 and 3 systems with nondiagonal linear part admitting a certain 
type of higher symmetry. We give master symmetries of systems determined to 
have only higher symmetries, and bi-Poisson formulation of those that turned 
out to possess conservation laws also.

\section{Second-order systems}\label{sec:2}

The class of two-component (1,1)-homogeneous of weight 2 systems with 
undetermined constant coefficients $c^i_j$ have the form
\begin{equation}\label{2ndord}
\left\{
\begin{array}{l}
\frac{du}{dt}=c^1_1u_{xx} + c^1_2v_{xx} + c^1_3 u u_{x} + c^1_4u v_{x} 
+ c^1_5 v u_{x}  + c^1_6 v v_{x} + c^1_7 u^3\\ 
\phantom{\frac{du}{dt}=}+ c^1_8 u^2 v + c^1_9 u v^2 + c^1_{10} v^3 \,,\\
\frac{dv}{dt}=c^2_1u_{xx} + c^2_2v_{xx} + c^2_3 u u_{x} + c^2_4 u v_{x} 
+ c^2_5 v u_{x}  + c^2_6 v v_{x} + c^2_7 u^3 \\
\phantom{\frac{dv}{dt}=}+ c^2_8 u^2 v + c^2_9 u v^2 + c^2_{10} v^3\,.
\end{array}\right . 
\end{equation}
The class of (1,1)-homogeneous of weight 3 systems with undetermined constant 
coefficients $l^i_j$ is
\begin{equation}\label{3rdord}
\left\{
\begin{array}{l}
\frac{du}{d\tau}=l^1_1u_{xxx}+ l^1_2 v_{xxx} + l^1_3uu_{xx} + l^1_4uv_{xx} 
+ l^1_5v u_{xx} + l^1_6v v_{xx} + l^1_7 u_{x}^2 + l^1_8 u_{x} v_{x}  \\
\phantom{u_{\tau}=}+ l^1_9 v_{x}^2+ l^1_{10} u^2 u_{x} + l^1_{11} u^2 v_{x} 
 + l^1_{12} v u u_{x} + l^1_{13} u v v_{x} + l^1_{14} v^2 u_{x} \\
\phantom{u_{\tau}=}+ l^1_{15} v^2 v_{x} + l^1_{16} u^4 + l^1_{17} u^3 v 
+ l^1_{18} u^2 v^2 + l^1_{19} u v^3 + l^1_{20} v^4 \,, \\
\frac{dv}{d\tau}= l^2_1u_{xxx}+l^2_2 v_{xxx} + l^2_3 uu_{xx} + l^2_4 uv_{xx} 
+ l^2_5 vu_{xx} + l^2_6 vv_{xx} + l^2_7 u_{x}^2+ l^2_8 u_{x}v_{x}   \\
\phantom{v_{t_3}=}+ l^2_9 v_{x}^2 + l^2_{10} u^2 u_{x} + l^2_{11} u^2 v_{x} 
+ l^2_{12}vuu_{x} + l^2_{13} u v v_{x} + l^2_{14} v^2 u_{x}  \\
\phantom{v_{t_3}=}+ l^2_{15} v^2 v_{x} + l^2_{16} u^4 + l^2_{17} u^3 v 
+ l^2_{18} u^2 v^2 + l^2_{19} u v^3 + l^2_{20} v^4 \,.
\end{array} \right .
\end{equation}
The class of (1,1)-homogeneous of weight 5 systems contains about 60 terms in 
each component. Therefore we omit writing this class explicitly.

Imposing compatibility condition \eqref{symcon} among the classes of systems 
\eqref{2ndord} and \eqref{3rdord}, 
we obtain a system of (initially bilinear) algebraic equations 
among the undetermined constants $c^i_j$ and $l^i_j$'s. Each solution 
of the system of constraints on constants $c^i_j$ and $l^i_j$ determines 
a system and its higher symmetry. 
The system of constraints are solved by the CRACK package \cite{WolCr}.
Among the systems obtained to possess a higher symmetry, the uncoupled ones and 
the triangular ones, i.e. those that reduce to successive scalar equations, 
are discarded as they are trivial. The remaining systems are searched for 
conservation laws by the package \cite{HR}.

Solutions of the compatibility condition \eqref{symcon} among the classes 
\eqref{2ndord} and \eqref{3rdord} with $c^1_1=c^2_2=1$, $c^2_1=0$, 
$c^1_2(=\epsilon)\neq0$, (i.e. for ${\bf A}={\bf J}_3^{(1,\epsilon)}$)
is given in the following proposition.
\begin{proposition}\label{prp2}
Any (1,1)-homogeneous of weight  2 system of form \eqref{2ndord} 
with  nondegenerate constant coefficient matrix of leading order terms having 
one dimensional eigenspace, possessing a symmetry from 
the (1,1)-homogeneous of weight  3 class of systems \eqref{3rdord}
is equivalent, 
by a linear change of variables $u$ and $v$ and rescaling of $x$ and $t$, 
to the following equation
\begin{equation}\label{2:bir}
\left\{
\begin{array}{l}
\frac{du}{dt}= B_2[u]+\epsilon\big(v_{x}+vu \big)_{x},\\
\frac{dv}{dt}= v_{xx} + 2uv_{x} -\epsilon\big(uv^2 + vv_{x} \big),
\end{array} \right .
\end{equation}
with $\epsilon\neq0$. 
\end{proposition}
\begin{remark}\label{rmr1} 
The characteristic of system \eqref{2:bir} appears like a linear combination 
of the characteristics of the triangular system 
\begin{equation}\label{2:iki}
\left\{
\begin{array}{l}
\frac{du}{dt}= B_2[u],\\
\frac{dv}{dt}= v_{xx} + 2uv_{x},
\end{array} \right .
\end{equation}
and the characteristic of the system 
\begin{equation}\label{2:uc}
\left\{
\begin{array}{l}
\frac{du}{dt}= v_{xx}+vu_{x} + uv_{x} \\
\frac{dv}{dt}=-uv^2 - vv_{x} ,
\end{array} \right .
\end{equation}
which has the degenerate nondiagonal matrix ${\bf A}={\bf J}_3^{(0,1)}$. 
Indeed, systems \eqref{2:bir}, \eqref{2:iki} and \eqref{2:uc} 
are all compatible with each other. However, despite the fact that these three 
systems have common symmetries (of higher orders),  each of these systems 
individually have different symmetry algebras: Some elements of the symmetry 
algebra of system \eqref{2:bir} are not admitted by systems \eqref{2:iki} and  
\eqref{2:uc} as a symmetry. Similarly, system \eqref{2:iki} (and \eqref{2:uc}) 
has its own symmetries not admitted by the the other two.
Therefore, the constant $\epsilon$ appearing like an arbitrary constant 
in a linear combination in the characteristic of system \eqref{2:bir} 
cannot be regarded as such, and $\epsilon\neq0$ is not a removable  
artifact of the formulation of the classification problem. 
\end{remark}

\begin{remark}
The Hopf-Cole transformation $u=\frac{\tilde u_{x}}{\tilde u}$ and  
$v=\frac{\tilde v}{\tilde u}$ not only linearizes system \eqref{2:bir} 
but makes it a triangular system.
\end{remark}
\begin{remark}
System \eqref{2:iki} becomes a first order equation of hydrodynamic type 
by $u=\tilde u_{x}$.
\end{remark}
\section{Third-order systems}
Setting
$l^1_1=l^2_2=1$, $l^2_1=0$,
$l^1_2(=\epsilon)\neq0$, (i.e. ${\bf A}={\bf J}_3^{(1,\epsilon)}$) 
in \eqref{3rdord}, and imposing compatibility condition \eqref{symcon} 
among the class of systems \eqref{3rdord} 
and the class of (1,1)-homogeneous of weight 5 systems,
we obtain the systems listed in the following proposition, 
as solutions of conditions imposed on the undetermined 
constants.
\begin{proposition}\label{prp3}
Any (1,1)-homogeneous of weight  3 system of form \eqref{3rdord} 
with a nondegenerate constant coefficient matrix of leading order terms having 
one dimensional eigenspace, possessing a symmetry from 
the (1,1)-homogeneous of weight 5 class of systems is equivalent, 
by a linear change of variables $u$ and $v$ and rescaling of $x$ and $t$, 
to one of the following eight equations:

\begin{equation}\label{3:1}
\left\{
\begin{array}{ll} 
\frac{du}{dt}=B_3[u]+\epsilon\big(v_{xx}  + 2uv_{x} + vu_{x}
+ u^2v \big)_{x} \,,\\
\frac{dv}{dt}=v_{xxx} + 3u_{x} v_{x} - 3vuu_{x} 
-\epsilon\big(vv_{xx} - v_{x}^2 + v^2u_{x} \big)\,,
\end{array} \right .
\end{equation}

\begin{equation}\label{3:2}
\left\{
\begin{array}{ll} 
\frac{du}{dt}=B_3[u] +\epsilon\big(v_{xx} + 2uv_{x} \big)_x\,,\\
\frac{dv}{dt}=v_{xxx} - 3uv_{xx}  + 3u^2v_{x} + 2\epsilon v_{x}^2 \,,
\end{array} \right .
\end{equation}

\begin{equation}\label{3:3}
\left\{
\begin{array}{ll} 
\frac{du}{dt}=B_3[u]+2\epsilon\big(v_{xx}  + 3uv_{x} + 2vu_{x}
+ 2u^2v\big)_{x} \,,\\
\frac{du}{dt}=v_{xxx} + 3uv_{xx} + 6u_{x} v_{x} + 3u^2v_{x}
-\epsilon\big( 4vv_{xx} - v_{x}^2 + 8v^2u_{x} + 8uvv_{x} + 4u^2v^2 \big)\,,
\end{array} \right .
\end{equation}

\begin{equation}\label{3:4}
\left\{
\begin{array}{ll}
\frac{du}{dt}=B_3[u]+\epsilon\big(v_{xx} + 2uv_{x}+vu_{x}  
+ u^2v\big)_{x}\,, \\
\frac{dv}{dt}=v_{xxx}  + 6u_{x}v_{x} 
-\epsilon\big(vv_{xx}- v_{x}^2 + v^2u_{x} \big)\,, 
\end{array} \right .
\end{equation}

\begin{equation}\label{3:5}
\left\{
\begin{array}{ll} 
\frac{du}{dt}=B_3[u]+3\epsilon\big(v_{xx} -vu_{x}\big)_{x} 
+3\epsilon^2\big(4vv_{xx}+3v_{x}^2-v^2u_{x}\big) 
+12\epsilon^3v^2v_{x}\,,\\
\frac{dv}{dt}= v_{xxx}-3(uv_{x}-u^2v)_{x}
+3\epsilon\big(3vv_{xx}+3v_{x}^2-2(v^2u)_{x}\big)
+21\epsilon^2v^2v_{x}\,,
\end{array} \right .
\end{equation}

\begin{equation}\label{3:6}
\left\{
\begin{array}{ll} 
\frac{du}{dt}=B_3[u]+\epsilon\big(v_{xx} + uv_{x} \big)_{x} \,,\\
\frac{dv}{dt}=v_{xxx} - 3uv_{xx} + 3v_{x}u_{x}  + 3 u^2 v_{x} 
+ 2\epsilon v_{x}^2 \,,
\end{array} \right .
\end{equation}

\begin{equation}\label{3:7}
\left\{
\begin{array}{ll} 
\frac{du}{dt}=B_3[u] +\epsilon\big(v_{xx} - uv_{x} -2vu_{x} 
- 2u^2v\big)_{x}\,,\\
\frac{dv}{dt}=v_{xxx}- 3(uv_{x}-u^2v)_{x} 
+\epsilon\big( 2vv_{xx} + v_{x}^2 - 6uvv_{x}  - 4v^2u_{x} \big)\,,
\end{array} \right .
\end{equation}

\begin{equation}\label{3:8}
\left\{
\begin{array}{ll} 
\frac{du}{dt}=B_3[u]+3\epsilon\big(2v_{xx} - 3v u_{x} - 4u^2v\big)_{x}
+3\epsilon^2\big( v^2u \big)_{x}\,,\\
\frac{dv}{dt}=v_{xxx} - 3(uv_{x} -u^2v)_{x} 
+3\epsilon \big(3vv_{x} - 4v^2u \big)_{x} 
+ 3\epsilon^2 v^2v_{x}\,,
\end{array} \right .
\end{equation}

where $\epsilon\neq0$.
\end{proposition}
None of the systems given in the above proposition is 
in the symmetry algebra of a second order system.   
Third order systems which are identified to be a symmetry of a second order 
system are discarded.

All systems \eqref{3:1}-\eqref{3:8} reduce to scalar third-order 
Burgers equation $\frac{du}{dt}=B_3[u]$, by $v=0$ reduction. 

Linear transformations of dependent variables preserve  homogeneity  of 
the classes considered but they are too restricted as an equivalence criterion. 
Therefore, hereafter we investigate each of the systems in the above proposition 
for a possible transformation, not necessarily invertible that can relate the 
system to a trivial one, to a known system, or to a nicer form of it; 
for a master symmetry, and for bi-Poisson formulation.

\subsection{System \eqref{3:1}}
System \eqref{3:1} transformed by $w=e^{\int u dx}$ and 
$z=e^{-\int u dx}(ve^{\int u dx})_x$ or 
equivalently in three stages $u=\frac{w_x}{w}, v=\frac{s}{w}, s_x=wz$,  
becomes the triangular equation
\begin{equation*}
\left\{
\begin{array}{ll} 
\frac{dw}{dt}=w_{xxx} + \epsilon(wz)_x\,, \\
\frac{dz}{dt}=z_{xxx} + \epsilon zz_x \,,
\end{array} \right .
\end{equation*}
which is trivial.

\subsection{System \eqref{3:2}} System \eqref{3:2} transformed by 
$u=\frac{p_{x}}{p}$ and $v_{x}=pq$, becomes
\begin{equation*}
\left\{
\begin{array}{ll} 
\frac{dp}{dt}=p_{xxx} + \epsilon(p^{2}q_{x}+3qpp_{x})\,, \\
\frac{dq}{dt}=q_{xxx} + \epsilon(q^{2}p_{x}+3pqq_{x})\,,
\end{array} \right .
\end{equation*}
which is the Sasa-Satsuma system \cite{SaSa}.
Poisson-symplectic formulation of the Sasa-Satsuma system is given 
by Sergyeyev and Demskoi in \cite{SergDems} 
with some misprints and by Wang in \cite{Wang1}.

\subsection{System \eqref{3:3}} 
System \eqref{3:3} has the master symmetry 
\begin{equation*}
{\bf M}^{\eqref{3:3}}=
\left(
\begin{array}{c} 
x\frac{du}{dt} + 2u_{xx} + 5uu_{x} + u^3 
+\epsilon\big( 3v_{xx} + 6vu_{x} + 8uv_{x} + 4u^2v\big) \\
x\frac{dv}{dt} + v_{xx} + 3vu_{x} -2\epsilon\big( vv_{x} + 2v^2u \big)
\end{array}
\right) .
\end{equation*}
So one can generate symmetries 
of arbitrarily high orders by using  
$ad_{\scriptscriptstyle {\bf M}^{\eqref{3:3}}}$ successively. 
Therefore  system \eqref{3:3}, 
obtained by acting $ad_{\scriptscriptstyle {\bf M}^{\eqref{3:3}}}$
on $x$-translation symmetry, is symmetry integrable.

Similar to the case in Remark~\ref{rmr1}, the coefficients 
of $\epsilon^0$ and $\epsilon^1$ in the characteristic of system~\eqref{3:3} 
and the characteristic of system~\eqref{3:3} as a whole, are all symmetries 
of each other. But their symmetry algebras do not overlap.

\subsection{System \eqref{3:4}} 
System \eqref{3:4} is a symmetry integrable system having the master symmetry 
\begin{equation*}
{\bf M}^{\eqref{3:4}}=
\left(
\begin{array}{c}
x\frac{du}{dt} + 2u_{xx} + 5uu_{x} + u^3 
+\frac13\epsilon\big(4v_{xx} +4vu_{x} +7uv_{x} +3u^2v\big)\\
x\frac{dv}{dt} + v_{xx} + 2uv_{x} -\frac13\epsilon\big(vv_{x} + uv^2\big)
\end{array}
\right).
\end{equation*}
Acted on $x$-translation symmetry,
$ad_{\scriptscriptstyle M^{\eqref{3:4}}}$ gives system~\eqref{3:4}.

\subsection{System \eqref{3:5}}
Symmetry integrable system \eqref{3:5} is obtained by acting 
$ad_{\scriptscriptstyle {\bf M}^{\eqref{3:5}}}$ to 
$x$-translation symmetry where the master symmetry is 
\begin{equation*}
{\bf M}^{\eqref{3:5}}=
\left(
\begin{array}{c} 
x\frac{du}{dt} + \frac52u_{xx} + 6uu_{x} + u^3 
+ \epsilon\big(6v_{xx} - 4vu_{x} -uv_{x}\big)
+\epsilon^2\big(18vv_{x}-v^2u\big) +4\epsilon^3 v^3 \\ 
x\frac{dv}{dt} + \frac32v_{xx}  -3uv_{x} + 3u^2v 
+\epsilon\big( 13vv_{x}-6v^2u\big) +7\epsilon^2 v^3 
\end{array}
\right).
\end{equation*}

\subsection{System \eqref{3:6}} Transforming system \eqref{3:6} by 
$v_x=z$ for convenience and setting without loss of generality $\epsilon=1$, 
we obtain 
\begin{equation}\label{3:5a}
\left\{
\begin{array}{l}
\frac{du}{dt}=B_3[u] + (z_{x} + uz)_x \,,\\
\frac{dz}{dt}=z_{xxx} + 3(u_{x}z-uz_{x}+u^{2}z)_{x} 
+ 4zz_{x},
\end{array}
\right .
\end{equation}
for which we have the following proposition.
\begin{proposition}
System~\eqref{3:5a} is bi-Poisson

\begin{equation*}
\frac{d}{dt}
\left(\!\!\!
\begin{array}{cc} 
u\\
z
\end{array}
\!\!\!\right)
=\begin{array}{c} 
{\mathcal H} \delta\tint z(u_{x}+\frac{1}{2}u^{2}+\frac{1}{3}z)\,dx
\end{array}
=\begin{array}{c} 
{\mathcal K} \delta\tint\frac{z}{2} \,dx\,, 
\end{array}
\end{equation*}
with the compatible pair of Poisson structures
\begin{equation*}
{\mathcal H}= \left(
\begin{array}{cc} 
-\frac{1}{3}D & D^{2}+2Du\\
-D^{2}+2uD & 4zD + 2z_{x}
\end{array}\right),\;\;\;
{\mathcal K}=\left(
\begin{array}{cc} 
{\mathcal K}_{11}&{\mathcal K}_{12}\\
-{\mathcal K}^{*}_{12}&{\mathcal K}_{22}
\end{array}\right),
\end{equation*}
where
\begin{equation*}
\begin{array}{l}
{\mathcal K}_{11}=-D^3 +u^2D+uu_{x} - u_{x} D^{-1} u_{x}  ,   \\
{\mathcal K}_{12}=D^{4}  + 4u D^{3}  + \big(9u_x +3z + 5u^2\big) D^{2} 
+ \big(7u_{xx}+5z_{x}+16uu_x+3uz+2u^3\big)D \\
\phantom{K_{12}=}+ \big(2u_{xxx} + 6uu_{xx} + 6u_{x}^{2} + 6u^{2}u_{x} 
+ 2z_{xx} + 3zu_{x} + 2uz_{x}\big) - u_{x} D^{-1} z_{x}, \\
{\mathcal K}_{22}=6zD^{3}+9z_{x}D^{2}
+\big(7z_{xx}+12u_{x}z-12z_{x}u+12u^{2}z+9z^{2}\big)D\\
\phantom{K_{22}=}+\big(2z_{xxx}-6uz_{xx}
+6zu_{xx}+12uu_{x}z+6u^{2}z_{x}+9zz_{x}\big)-z_{x}D^{-1}z_{x}.
\end{array}
\end{equation*}
Here, $D^{-1}$ denotes the formal inverse of $D$ and ${\mathcal K}^{*}_{12}$ 
is the formal adjoint of ${\mathcal K}_{12}$. 
\end{proposition}

Skew-adjoint local operator ${\mathcal H}$ satisfies Jacobi identity.
Therefore it is a Poisson structure \cite{BDSK09}. 
To show that the skew-adjoint nonlocal operator ${\mathcal K}$ 
is a Poisson structure and  compatible with ${\mathcal H}$, 
it suffices to verify that the fractional decomposition of nonlocal operator 
$\alpha {\mathcal H}+{\mathcal K}={\mathcal A}{\mathcal B}^{-1}$, 
by  local operators 
${\mathcal A}=(\alpha {\mathcal H}+{\mathcal K}){\mathcal B}$ 
and
\begin{equation*}
{\mathcal B}=\left(
\begin{array}{cc}
\frac{1}{u_x}D & 0 \\
0 & \frac{1}{z_x}D\\
\end{array}\right)
\end{equation*}
satisfies  the Jacobi identity 
in terms of operators 
${\mathcal A}$ and ${\mathcal B}$ in \cite{DSK13} 
regardless of the value of constant $\alpha$. 
The case $\alpha=0$ corresponds to Jacobi identity for 
${\mathcal K}$ and $\alpha\neq0$ 
to the compatibility of ${\mathcal H}$ and ${\mathcal K}$. 
These completely  algorithmic computations which does not 
require any user intervention, are carried out by computer.

Recall that the case of arbitrary nonzero $\epsilon$ can be 
recovered simply by transforming all above quantities by 
$z\rightarrow \epsilon z$.

\subsection{System \eqref{3:7}}
As in the previous system, for the system \eqref{3:7} with $\epsilon=1$,
 we have the following proposition.
\begin{proposition}
System \eqref{3:7} is bi-Poisson
\begin{equation*}
\frac{d}{dt}
\left(\!\!\!
\begin{array}{cc} 
u\\
v
\end{array}
\!\!\!\right)
=\begin{array}{c} 
{\mathcal H} \delta\tint 
\big(-u_xv_x+3vuu_x-\frac23v^2u_x-\frac13v_x^2 +vu^3-\frac43u^2v^2\big)\,dx
\end{array}
=\begin{array}{c} 
{\mathcal K} \delta\tint uv\,dx
\end{array}
\end{equation*}
with the compatible pair of Poisson structures 
\begin{equation*}
{\mathcal H}
=\left(
\begin{array}{cc} 
\frac13D & D - \frac23 D \circ v(D+2u)^{-1}\\
D +\frac23 (D-2u)^{-1} \circ vD 
&-\frac23\big((D-2u)^{-1} v^2 + v^2 (D+2u)^{-1}\big) 
\end{array}\right),
\end{equation*}
and ${\mathcal K}$ having components
\begin{equation*}
\begin{array}{l}
{\mathcal K}_{11}=D^3 - u^{2} D - uu_{x} + u_{x} D^{-1}u_{x}  ,   \\
{\mathcal K}_{12}=D^3 + 2 (u -v) D^2 + ( 3u_{x}  - 3v_{x}  - 2uv + u^2)D \\ 
\phantom{K_{12}=}+ ( u_{xx} - v_{xx} +2uu_{x}-2vu_{x} - uv_{x} ) 
+ u_{x}  D^{-1} v_{x}, \\ 
{\mathcal K}_{21}=-{\mathcal K}_{12}^{*},\\
{\mathcal K}_{22}=-D ( v_{x}   - 2uv + 2v^2 )-(v_{x}- 2uv + 2v^2)D
+v_{x}D^{-1}v_{x}. 
\end{array}
\end{equation*}
\end{proposition}
A fractional decomposition
${\mathcal K}={\mathcal A}_{\mathcal K}({\mathcal B}_{\mathcal K})^{-1}$ 
of nonlocal skew-adjoint operator
${\mathcal K}$ is obtained by the local operators 
\begin{equation*}
\mathcal B_{\mathcal K}=\left(
\begin{array}{cc}
\frac{1}{u_x}D & 0\\
0 & \frac{1}{v_x}D\\
\end{array}\right), 
\end{equation*}
and ${\mathcal A}_{\mathcal K}={\mathcal K}{\mathcal B}_{\mathcal K}$.
Skew-adjoint operator ${\mathcal K}$ satisfies the Jacobi identity 
in terms of the local operators 
${\mathcal A}_{\mathcal K}$ and ${\mathcal B}_{\mathcal K}$ in \cite{DSK13}. 
Therefore ${\mathcal K}$ is a Poisson structure. A fractional decomposition 
for the skew-adjoint operator ${\mathcal H}$ is obtained as 
${\mathcal H}={\mathcal A}_{\mathcal H}({\mathcal B}_{\mathcal H})^{-1}$ 
by the local operators 
\begin{equation*}
{\mathcal A}_{\mathcal H}
=\left(
\begin{array}{cc}
 \frac13 D \circ vR\big(D-2u\big)&\big(D^2+2D\circ(u-\frac{v}{3})\big)
\circ R\big(D-2u\big) \\
\big(D+\frac23 v\big)\circ vR\big(D-2u\big)-\frac23 & 
-\frac{4}{3}\big(v^2R(D-2u)-1\big)\\
\end{array}\right), 
\end{equation*}
\begin{equation*}
{\mathcal B}_{\mathcal H}
=\left(
\begin{array}{cc}
 vR\big(D-2u\big)& 0\\
0 & (D+2u)\circ R(D-2u)\\
\end{array}\right), 
\end{equation*}
where $R=\frac{1}{vv_x-2uv^2}$.
Local operators ${\mathcal A}_{\mathcal H}$ and ${\mathcal B}_{\mathcal H}$ 
satisfies Jacobi identity for ${\mathcal H}$ \cite{DSK13}. 
Therefore ${\mathcal H}$ is a Poisson structure. For the compatibility of 
${\mathcal H}$ and ${\mathcal K}$ one needs to 
verify the Jacobi identity for ${\mathcal H}+{\mathcal K}$. 
For ${\mathcal H}+{\mathcal K}$ we have a fractional  decomposition 
${\mathcal H}+{\mathcal K}={\mathcal A}_{{\mathcal H}+{\mathcal K}}
({\mathcal B}_{{\mathcal H}+{\mathcal K}})^{-1}$ by local operators 
${\mathcal A}_{{\mathcal H}+{\mathcal K}}=({\mathcal A}_{\mathcal H}+
{\mathcal K}{\mathcal B}_{\mathcal H}){\mathcal S}$, 
${\mathcal B}_{{\mathcal H}+{\mathcal K}}
={\mathcal B}_{\mathcal H}{\mathcal S}$ with
\begin{equation*}
{\mathcal S}
=\left(
\begin{array}{cc}
\frac{-1}{F}D & 0\\
0 & \frac{1}{G}D\\
\end{array}\right),
\end{equation*}
where $F= \big(vu_xR\big)_{x} + 2vuu_{x}R$ and 
$G=\big(v_{xx}R\big)_{x} - 2v_{x}\big(uR)_{x} - 4u^2v_{x}R$.
The Jacobi identity for ${\mathcal H}+{\mathcal K}$, in terms of 
${\mathcal A}_{{\mathcal H}+{\mathcal K}}$ and 
${\mathcal B}_{{\mathcal H}+{\mathcal K}}$ is verified to be satisfied. 
So Poisson structures ${\mathcal H}$ and ${\mathcal K}$ are compatible.

System \eqref{3:7}, transformd by $u=\frac{w_{xx}}{w_{x}}$ and  $v=w_xz_x$ 
becomes 
\begin{equation*}
\left\{
\begin{array}{ll} 
\frac{dw}{dt}=w_{xxx} + \epsilon w_x(w_xz_{xx}-w_{xx}z_x)\,, \\
\frac{dz}{dt}=z_{xxx} + \epsilon z_x(w_xz_{xx}-w_{xx}z_x)\,,
\end{array} \right . 
\end{equation*}
which can be complexified through $\psi=w+iz$ as 
\begin{equation*}
\begin{array}{l} 
\frac{d\psi}{dt}=\psi_{xxx}+\frac{\epsilon}{2i}
\psi_x\big(\psi^{*}_{x}\psi_{xx}-\psi_{x}\psi^{*}_{xx}\big).
\end{array} 
\end{equation*}
This equation is the potential form of an equation given with  
its Lax pairs by Tsuchida in \cite{Tsc10}.

\subsection{System \eqref{3:8}} 
For the convenience of Poisson structures, besides taking $\epsilon=1$ 
in system~\eqref{3:8}, we transform it by $w=\frac12(u+v)$, $z=\frac12(u-v)$, 
and $\tau=-t$. As a result we get the system
\begin{equation}\label{3:8a}
\left\{
\begin{array}{l}
\frac{dw}{d\tau}=2 w_{xxx} - 3 z_{xxx}- 12 wz_{xx}+ 6 zz_{xx} 
- 12 w_{x} z_{x}  - 48 w^2 w_{x}   
+ 12 z^2w_{x} + 24 wzz_{x} + 6 z_{x}^2\,,\\
\frac{dz}{d\tau}=3 w_{xxx} - 4 z_{xxx} - 12 ww_{xx} + 6 zw_{xx} 
- 12 w_{x}^2 + 6 w_{x} z_{x} 
- 24 zww_{x} - 12 w^2z_{x} + 24 z^2z_{x}\,,
\end{array}
\right .
\end{equation}
for which we have the following proposition. 
\begin{proposition}
System~\eqref{3:8a} is bi-Poisson
\begin{equation*}
\begin{array}{ll}
\frac{d}{d\tau}
\left(\!\!\!
\begin{array}{c} 
w\\
z
\end{array}
\!\!\!\right)
&={\mathcal H} \delta\tint \big(- w_{x}^2
+ 3 w_{x} z_{x} - 2 z_{x}^2
-6 w^2 z_{x} - 3 z^2 w_{x} 
- 4 w^4 + 6 w^2 z^2 - 2 z^4
\big)\,dx \\
&={\mathcal K}\delta\tint \frac12\big(w^2-z^2\big)\,dx \,,
\end{array}
\end{equation*}
with the compatible pair of Poisson structures
\begin{equation*}
{\mathcal H}= \left(
\begin{array}{cc} 
 D & 0\\
 0 & -D
\end{array}\right)\,,
\end{equation*}
and ${\mathcal K}$ having elements
\begin{equation*}
\begin{array}{l}
{\mathcal K}_{11}=\big(2A_1 + 2D^{2}\circ A_2 + \frac{1}{2} D^{4} \big)
\circ\big( D + 4w \big)^{-1} 
+ ( D - 4w )^{-1} \circ\big(2A_1 + 2A_2 D^{2} + \frac{1}{2} D^{4} \big) \\
\phantom{K_{11}=}+12 w_{x} D^{-1}w_{x}  \,,  \\
{\mathcal K}_{12}=\big( D - 4w \big)^{-1}\circ\big(12B_1+2B_2 D+2B_3 D^{2}  
-4z D^{3} + 2 D^{4} \big)  
+ 12 w_{x} D^{-1}z_{x} \,,\\
{\mathcal K}_{21}=-{\mathcal K}^{*}_{12}\,,\\
{\mathcal K}_{22}= \big( 3C_1  +   6 D^{2}\circ C_2 + \frac{3}{2} D^{4} \big)
\circ\big(D + 2z \big)^{-1} 
+ \big(D - 2z\big)^{-1}\circ \big(3C_1 + 6C_2 D^{2}+\frac{3}{2} D^{4}\big)\\
\phantom{K_{22}=}+12 z_{x}D^{-1}z_{x} \,,\\
\end{array}
\end{equation*}
where 
\begin{equation*}
\begin{array}{ll}
A_1 =  - w_{xxx} + z_{xxx} + 24 ww_{xx}- 12 wz_{xx} - 2 zz_{xx} + 16 w_{x}^{2}
- 6 z_{x}^{2}+ 24  w^{2}z_{x} \\
\phantom{2A=}- 156 w^2 w_{x}  + 24 wzz_{x} + 8 z^2z_{x} + 108 w^{4} 
- 24 w^{2} z^{2} - 4 z^{4} \,, \\
A_2=4 w_{x} - 3 z_{x} - 12 w^{2} + 3 z^{2} \,, \\
B_1=- zw_{xx} - w_{x} z_{x} + 4 zww_{x}\,,\\
B_2= 3 w_{xx} - 2 z_{xx}  - 12 ww_{x} - 18 zw_{x} + 4 zz_{x} + 36 w^{2} z 
- 4 z^{3} \,, \\
B_3= 9 w_{x} - 6 z_{x} - 18 w^{2} + 2 z^{2} \,, \\
C_1=- z_{xxx} + 8 zz_{xx} + 6 z_{x}^{2} - 24 z^{2}z_{x} + 8 z^{4}\,,  \\
C_2=z_{x} - 2 z^{2}\,.  \\ 
\end{array}
\end{equation*}
\end{proposition}
It is obvious that  ${\mathcal H}$ is a Poisson structure. 
The operator ${\mathcal K}$ is also a Poisson structure compatible with 
${\mathcal H}$ but fractional decomposition of ${\mathcal K}$ 
is a long expression to write explicitly. 

Transforming  system~\eqref{3:8} by $u=\frac{r_x}{r}$ and $v=rs$, we obtain
\begin{equation*}
\left\{
\begin{array}{ll} 
\frac{dr}{dt}=r_{xxx}+3\epsilon \big(2r^2s_{xx}+4rr_xs_x-sr_x^2-srr_{xx} \big)
+3\epsilon^2s^2r^2r_x\,, \\
\frac{ds}{dt}=s_{xxx}+3\epsilon r\big(ss_{xx}+3s_x^2 \big) 
+3\epsilon^2r^2s^2s_x\,,
\end{array} \right .
\end{equation*}
which is again a polynomial system. 
In this final form of system~\eqref{3:8}, the coefficient of leading order terms 
is the identity matrix ${\bf A}={\bf J}_1$ and the parameter $\epsilon$ 
is a factor of all nonlinear terms.

\section{Conclussion}

With the case of nondiagonal coefficient matrix of leading order linear terms, 
higher symmetry  classification of $(1,1)$-homogeneous of 
weight 2 and 3 classes of two-component systems are completed. 

Although we naively completed
our classification with nondiagonal matrix of leading order terms, 
all the systems obtained having a higher symmetry turned out to be systems 
admitting  the Hopf-Cole transformation $u=\ln(w)_x$
after which the coefficient matrix of leading order terms becomes 
the identity matrix ${\bf J}_1$. 
By admission of a differential substitution, we mean that the system  
remains to be a system of differential (but not integro-differential) 
functions, after the substitution. 

On the other hand, the parameter $\epsilon$ turned out to be essential 
for all systems obtained to be nontrivial: 
Although it is possible to set $\epsilon=0$ in all systems obtained 
($\epsilon$ is nowhere in denominator), 
this reduction of all systems is triangular.

Leaving aside the system~\eqref{3:2} which is determined to be related to the
well known Sasa-Satsuma system,
we obtained three third-order symmetry integrable systems 
\eqref{3:3}-\eqref{3:5} and their master symmetries. 
These systems appear to be new. However, as C-integrable systems in the 
terminology of Calogero, these symmetry integrable systems are presumably 
useful nonlinear forms of linear systems. 

The second of the last three bi-Poisson systems \eqref{3:6}-\eqref{3:8} 
is related to a system given by Tsuchida. The remainin systems 
\eqref{3:6} and \eqref{3:8} seem to be new.
The Lenard-Magri schemes of these bi-Poisson  systems are verified 
to work for the next two members of the symmetry hierarchy.
Therefore these systems are  integrable in the sense of 
infinite conservation laws.

\section*{Acknowledgements}

DT was supported by TUBITAK PhD Fellowship for Foreign Citizens.

\end{document}